# Aligning Heterogeneous DFT Datasets: A Graph Neural Network Approach to Cross-Functional Formation Energies

*Yidong Huang,*[1,2] *Tenglong Lu,*[1,3] *Hanwen Kang,*[1,2] *Junfeng Huang,*[1,2]
*Sheng Meng,* [1*] *Miao Liu,* [1,3*]

[1] Beijing National Laboratory for Condensed Matter Physics, Institute of Physics, Chinese Academy of Sciences, Beijing 100190, China;

[2] University of Chinese Academy of Sciences, Beijing 100049, China

[3] Dongguan Institute of Materials Science and Technology, Dongguan, Guangdong 523808, China

* E-mail: smeng@iphy.ac.cn, mliu@iphy.ac.cn

## Abstract

Heterogeneous density functional theory (DFT) calculations, particularly plane-wave implementations, introduce systematic formation energy errors ranging from tens to hundreds of meV/atom, depending on the selection of exchange-correlation functionals, kinetic energy cutoffs, pseudopotentials, and dispersion corrections. As demonstrated by the MatPES dataset, identical structures can exhibit an average energy discrepancy of 107 meV/atom between PBE and r$^2$SCAN calculations. Such method-dependent discrepancies hinder the integration of multi-source DFT data, greatly limiting the scale and quality of datasets for training robust materials AI models. Here, we resolve this fundamental data silo barrier via graph-based transfer learning. Leveraging 380,190 structurally paired PBE–r$^2$SCAN entries from the MatPES database, we train a structure-aware graph neural network to predict cross-functional energy residuals and align inconsistent DFT energy scales. By adopting GPTFF model architecture, the model converts conventional PBE energies to r$^2$SCAN-level accuracy with a mean absolute error of 14.3 meV/atom, compared with 18.2 meV/atom achieved by CHGNet. This versatile approach effectively upgrades massive legacy PBE datasets to high-precision r$^2$SCAN standards. It enables reliable predictions of phase stability, battery voltage profiles, and reaction thermodynamics, while allowing the integration of multi-source DFT data to advance the development of high-performance materials foundation models.

# 1 Introduction

Recent advances in high-throughput first-principles calculations, artificial intelligence, and automated experimentation are reshaping the paradigm of materials research, shifting materials discovery from experience-driven trial and error toward data-driven design [1-5]. In this context, large, high-quality, and standardized materials databases play a central role. Databases such as the Materials Project, OQMD, AFLOW, JARVIS, and Atomly collect crystal structures, chemical compositions, total energies, formation energies, electronic structures, and thermodynamic stability data, thereby providing essential inputs for high-throughput screening, phase-diagram construction, battery-materials design, reaction thermodynamics, and machine-learning model development [6-10]. Nevertheless, for key energetic properties such as formation energies, phase stability, and reaction energies, sheer data volume is insufficient. The energetic consistency and cross-dataset comparability of computational results are equally critical for robust data integration and reliable downstream modeling.

In practical materials research, no single data source can simultaneously deliver large data volumes, broad chemical space coverage, and high energetic accuracy. Large datasets generated via low-cost GGA/PBE functionals [11] feature wide chemical diversity and have been widely used to train universal machine learning interatomic potentials and property prediction models [12-17]. In comparison, higher-level functionals such as $r^2$SCAN yield more accurate descriptions of formation energies, phase stability, and solid-state potential energy surfaces but incur substantially higher computational costs, resulting in smaller dataset sizes [18-21]. Building a rigorous energy mapping scheme across different theoretical levels enables the integration of the scale advantage of low-cost calculations and the accuracy advantage of high-fidelity computations. This synergy can substantially strengthen the reliability of materials screening, interatomic potential development, phase stability evaluation, and reaction energy prediction [22,23,28-30].

Notably, multi-source datasets cannot be directly merged for immediate use. Different computational datasets vary substantially in key calculation parameters, including exchange-correlation functionals, pseudopotentials, plane-wave energy cutoffs, k-point sampling, magnetic state treatments, structural relaxation protocols, and elemental reference states. These methodological discrepancies introduce systematic, source-dependent offsets in formation energy labels, obscuring the intrinsic composition–structure–energy correlations that machine learning models aim to learn [6-9,22-25]. More critically, even minor systematic errors can propagate to erroneous evaluations of phase stability, convex-hull stability, battery voltage profiles, and chemical reaction driving forces, ultimately undermining model generalization performance [22,23,32,38]. Thus, although multi-source materials data fusion can substantially expand overall dataset size, its core challenge resides in the reliable elimination and alignment of energy offsets across heterogeneous computational datasets.

To address such energy inconsistencies, multiple methodological strategies have been proposed to unify DFT energy scales, including GGA/GGA+U compatibility schemes, fitted elemental-phase reference energies (FERE), coordination-corrected enthalpies (CCE), and PBE(+U)–r$^2$SCAN formation-energy mixing [22,24-27]. These existing methods improve cross-dataset energetic comparability by optimizing internal database compatibility, calibrating DFT-computed formation enthalpies against experimental benchmarks, and unifying reference states across different functionals. Nevertheless, these conventional correction methods often rely on relatively simple regression or reference-based correction schemes, which can limit their generalizability and correction accuracy. They are usually designed for specific datasets or correction protocols and may fail to characterize the complex, structure-dependent nature of cross-functional energy deviations. Recent cross-functional transfer-learning and multifidelity learning studies further suggest that such functional-dependent energy deviations can be learned from data-driven models [28-30]. Moreover, cross-functional energy biases are highly sensitive to material composition, local coordination, and bonding environments—structural features that can be inherently captured by machine-learning interatomic potential (MLIP) architectures, which are specifically designed to resolve local structure-dependent energy variations [12,14-17].

Therefore, in this work, we design a neural network framework based on mature MLIP architectures, adopting GPTFF and CHGNet as representative backbone models to learn and predict structure-dependent energy residuals between PBE and r$^2$SCAN calculations of inorganic materials. The trained model implements structure-aware corrections to raw PBE formation energies, calibrating them to the r$^2$SCAN energy scale. This correction strategy enables r$^2$SCAN-level accuracy in phase diagram construction, battery voltage evaluation, and chemical reaction energy prediction. Although we demonstrate and validate our alignment approach using GPTFF and CHGNet on the PBE–r$^2$SCAN dataset pair, our method is not limited to these specific models or functional combinations. It features excellent generalizability and can be readily extended to diverse MLIP frameworks and heterogeneous DFT datasets with varied computational configurations, providing a universal solution for multi-source materials data alignment and fusion.

# 2 Methods

## 2.1 Dataset preparation and learning target

Structurally paired PBE and r$^2$SCAN calculations were obtained from the MatPES dataset [21]. Per-atom formation energies were evaluated with pymatgen [31] using elemental reference states defined separately for each functional. For each functional, the lowest-energy unary phase available in the corresponding dataset was independently selected as the elemental reference state. Thus, PBE and r$^2$SCAN formation energies were calculated using their respective lowest-energy elemental phases.

Only entries with identical compositions, complete energies, and unambiguous one-to-one structural correspondence were retained, yielding 380,190 paired structures. The learning target was the per-atom r$^2$SCAN-minus-PBE formation-energy difference. Model-corrected energies were obtained by adding the predicted difference to the original PBE formation energy.

## 2.2 Models and training

GPTFF [14] and CHGNet [12] were used as graph-neural-network backbones, with crystal structures as inputs and cross-functional formation-energy differences as outputs. The paired dataset was randomly divided into training, validation, and test sets in an 80:10:10 ratio. Models were selected on the validation set and evaluated on the held-out test set using the mean absolute error.

## 2.3 Thermodynamic validation

Binary phase diagrams for Fe-O, Mn-O, Ti-Cl, Mo-Cl, V-S, and Cu-S were constructed with pymatgen [31]. Each convex hull was evaluated from the full candidate-structure set using PBE, r$^2$SCAN, or model-corrected formation energies. For visualization, the plotted compositions were defined as the union of the convex-hull vertices obtained from the three energy scales; at each selected composition, the lowest-energy structure under each method was shown. Stable phases and energies above hull were then compared across the three energy scales.

Average Li deintercalation voltages were evaluated for $LiCoO_2$ and $LiMn_2O_4$, with $LiFePO_4$ results reported in the Supplementary Information. Candidate Li/vacancy orderings were generated using pymatgen, relaxed at the PBE level, and evaluated with PBE and r$^2$SCAN static calculations following the MatPES computational settings [21]. The lowest-energy configuration at each Li content was retained, and average voltages between adjacent compositions were calculated from reaction energies using the established first-principles procedure [36,37]. The same workflow was applied to the model-corrected energies.

Solid-state reaction pathways were generated and balanced with pymatgen reaction tools [31]. Reaction energies were obtained from the stoichiometrically weighted formation energies and compared using PBE, r$^2$SCAN, model-corrected, and experimental thermochemical data [39,40].

# 3 Results

## 3.1 PBE-r$^2$SCAN formation-energy differences and chemical dependence

We first compare PBE and r$^2$SCAN formation energies against experimental formation enthalpies to assess the reliability of the two functionals for thermodynamic stability prediction. From the overlap between the Materials Project experimental formation-energy data and the dataset constructed in this work, 1,234 shared compounds are obtained [23]. As shown in Fig.1(a), The PBE formation energies show a mean absolute error of 164 meV/atom relative to experiment, whereas the r$^2$SCAN MAE decreases to 87 meV/atom. This confirms that r$^2$SCAN provides a clear improvement over PBE for formation-energy prediction and supports the physical and practical value of mapping large-scale PBE formation energies toward the r$^2$SCAN scale.

We next construct a structurally paired PBE-r$^2$SCAN formation-energy dataset from MatPES [21]. The two functionals share 380,190 structurally matched samples. PBE and r$^2$SCAN formation energies are strongly correlated overall, but substantial deviations remain. As shown in Fig.1(b) and Fig.1(d), The mean absolute PBE-r$^2$SCAN difference is 107 meV/atom, and the maximum difference reaches 1.960 eV/atom. These deviations are not random noise but systematic offsets with learnable structure, motivating the use of their formation-energy difference ($\Delta E_f$) as the machine-learning target. Learning this difference allows the large-scale coverage of PBE data to be retained while introducing r$^2$SCAN-level energetic fidelity.

To further understand the origin of the PBE-r$^2$SCAN differences, we empirically classify compounds according to elemental type and Pauling electronegativity difference into ionic inorganic materials, covalent inorganic materials, metallic-bonded materials, and mixed-bonding materials. As shown in Fig.1(c), Ionic compounds show the largest PBE-r$^2$SCAN formation-energy differences, with a mean absolute difference of 181 meV/atom. In comparison, the mean differences for covalent, metallic, and mixed-bonding materials are 62, 73, and 86 meV/atom, respectively.

These results demonstrate a pronounced chemical dependence of the PBE-r$^2$SCAN bias. In ionic compounds, different functionals are expected to differ more strongly in their descriptions of charge transfer, localized electronic states, and cation-anion interactions, leading to a larger need for formation-energy correction. The result also indicates that a simple global linear shift cannot fully remove the PBE-r$^2$SCAN discrepancy; models must account for both elemental composition and local structural environment.

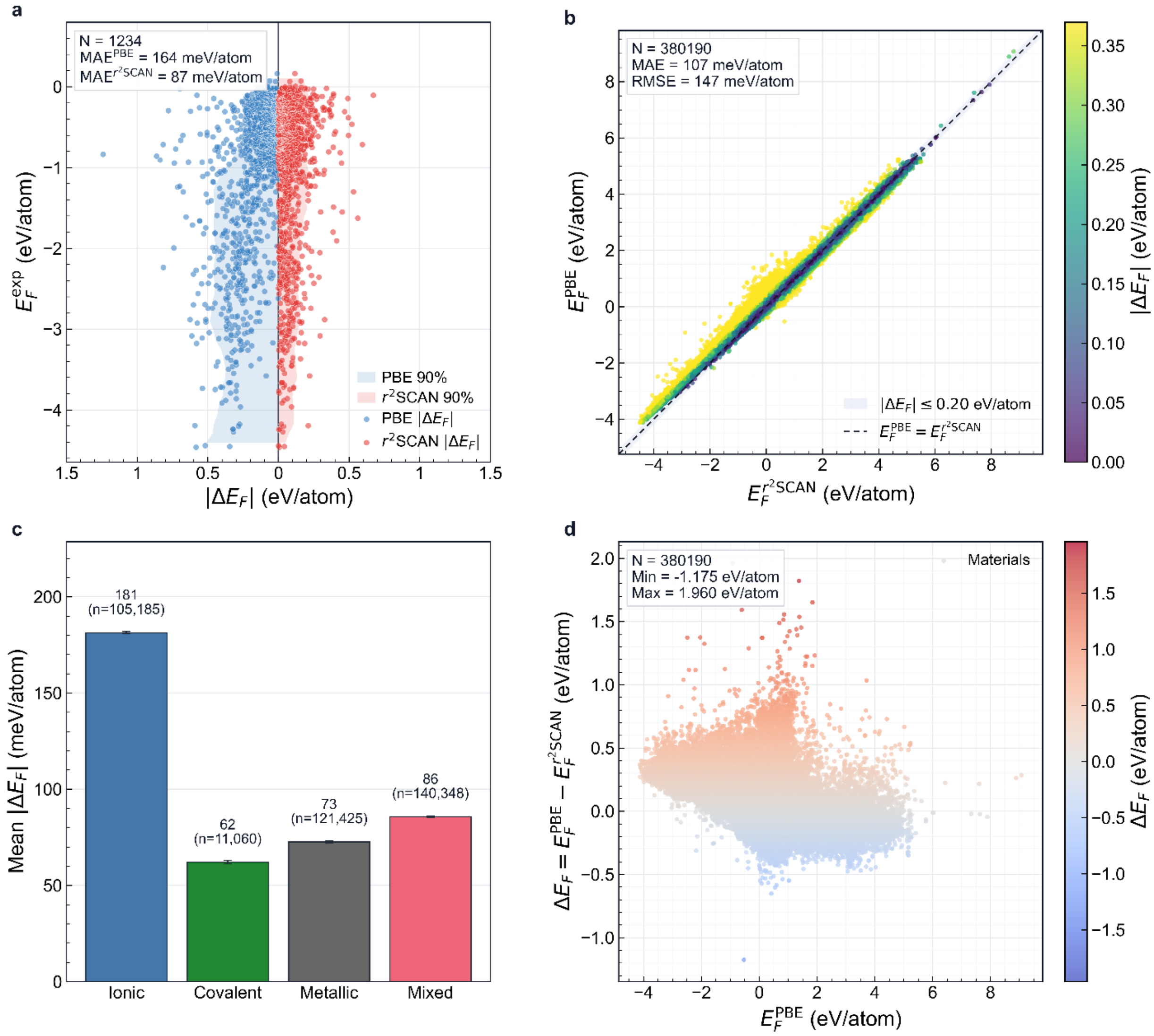

Figure 1. Differences between the PBE and r$^2$SCAN formation-energy datasets and their chemical origin. (a) Absolute errors of PBE and r$^2$SCAN formation energies relative to experimental formation enthalpies. (b) Correlation between paired PBE and r$^2$SCAN per-atom formation energies; color indicates the absolute PBE-r$^2$SCAN formation-energy difference. (c) Mean absolute PBE-r$^2$SCAN formation-energy difference for different bonding categories. (d) Relationship between PBE formation energy and the absolute cross-functional formation-energy difference.

### 3.2 Test-set performance of cross-functional formation-energy-difference prediction

To map PBE formation energies toward the r$^2$SCAN energy scale, we construct a formation-energy-difference dataset using the per-atom formation-energy difference ($\Delta E_f$) of the same structure as the supervised label. We train two open-source graph-neural-network materials models, GPTFF and CHGNet, on this dataset with 80% of the data used for training, 10% for validation, and 10% for testing. Figures 2a and 2b show the test results for GPTFF and CHGNet, respectively. GPTFF achieves an MAE of 14.3 meV/atom, while CHGNet achieves an MAE of 18.2 meV/atom. We therefore use the GPTFF model for subsequent correction and analysis. The trained model is then applied to predict formation-energy differences for structures in the PBE dataset, and the predicted residuals are added to the PBE formation energies to obtain the model-corrected formation-energy dataset.

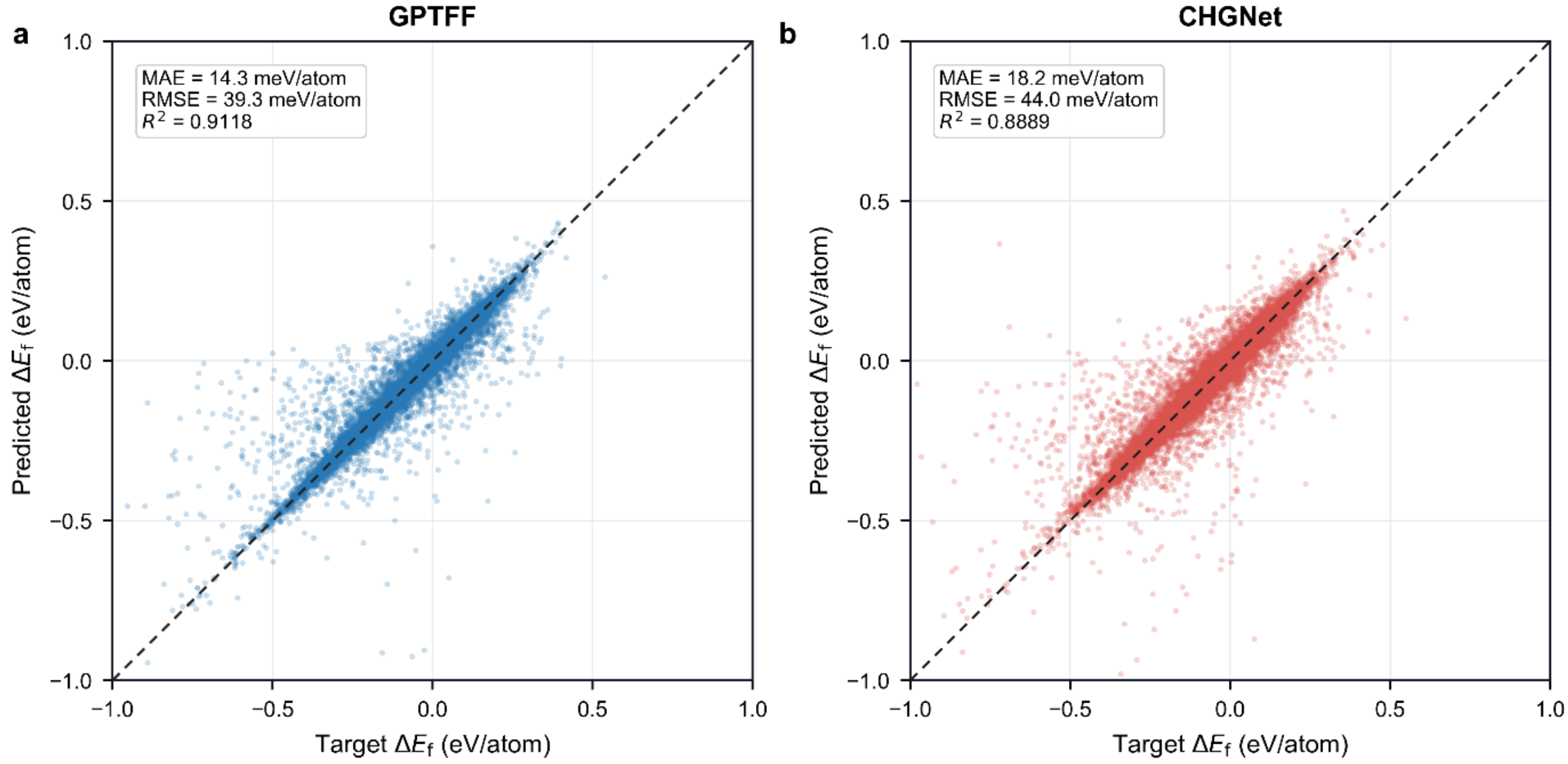

Figure 2. Test-set prediction performance of GPTFF and CHGNet for cross-functional formation-energy differences. (a) Parity plot of GPTFF predictions versus target values. (b) Parity plot of CHGNet predictions versus target values. The dashed line indicates the ideal y = x relationship; inset boxes report MAE, RMSE, and $R^2$.

### 3.3 Phase-stability validation

Although test-set MAE measures pointwise predictive accuracy, convex-hull stability is controlled by the relative energy ordering among competing phases. We therefore constructed binary phase diagrams for Fe-O, Mn-O, Ti-Cl, Mo-Cl, V-S, and Cu-S using PBE, r$^2$SCAN, and GPTFF-corrected formation energies (Figure 3). The displayed compositions correspond to the union of the convex-hull vertices predicted by the three methods. At each composition, the lowest-energy structure under each energy scale is plotted; filled symbols denote phases on the corresponding convex hull, whereas open symbols denote the lowest-energy structure at that composition that remains above the hull.

The correction recovers several $r^2$SCAN-stable phases that are placed above the PBE hull. $Fe_2O_3$, MnO, and $TiCl_2$ lie 16.7, 38.7, and 27.2 meV/atom above the PBE hull, respectively, but are stable in both $r^2$SCAN and GPTFF-corrected phase diagrams. Similar recovery is obtained for $MoCl_3$, $V_5S_8$, and $Cu_2S$, whose PBE energies above hull are 48.9, 13.9, and 9.2 meV/atom, respectively. Overall, GPTFF reproduces the $r^2$SCAN stability classification for six of the eight phases whose PBE and $r^2$SCAN classifications differ, corresponding to a recovery rate of 75%.

The $r^2$SCAN convex-hull topology is nevertheless not reproduced exactly. $FeO_2$, $Mn_5O_8$, and $V_7S_8$ are stable in $r^2$SCAN but remain 21.5, 64.8, and 86.9 meV/atom above the GPTFF-corrected hull, respectively. Conversely, $MoCl_4$, $V_3S_5$, and $CuS_2$ are predicted to lie on the corrected hull although their $r^2$SCAN energies above hull are 115.9, 13.2, and 4.0 meV/atom, respectively. The latter two cases are close to the thermodynamic stability boundary, whereas the $MoCl_4$ result indicates a more substantial local phase-ordering error. Because energy above hull is determined jointly by a phase and its competing decomposition products, these discrepancies cannot be attributed to the target phase alone without analyzing the associated decomposition reactions.

Overall, GPTFF correction substantially improves the PBE thermodynamic ordering and recovers the principal features of the $r^2$SCAN convex hulls, but a low average formation-energy error does not guarantee an identical set of stable phases. For phase-diagram screening, predictions within several tens of meV/atom of the hull should therefore be interpreted as stability-sensitive and, when decisive, verified by explicit high-fidelity calculations.

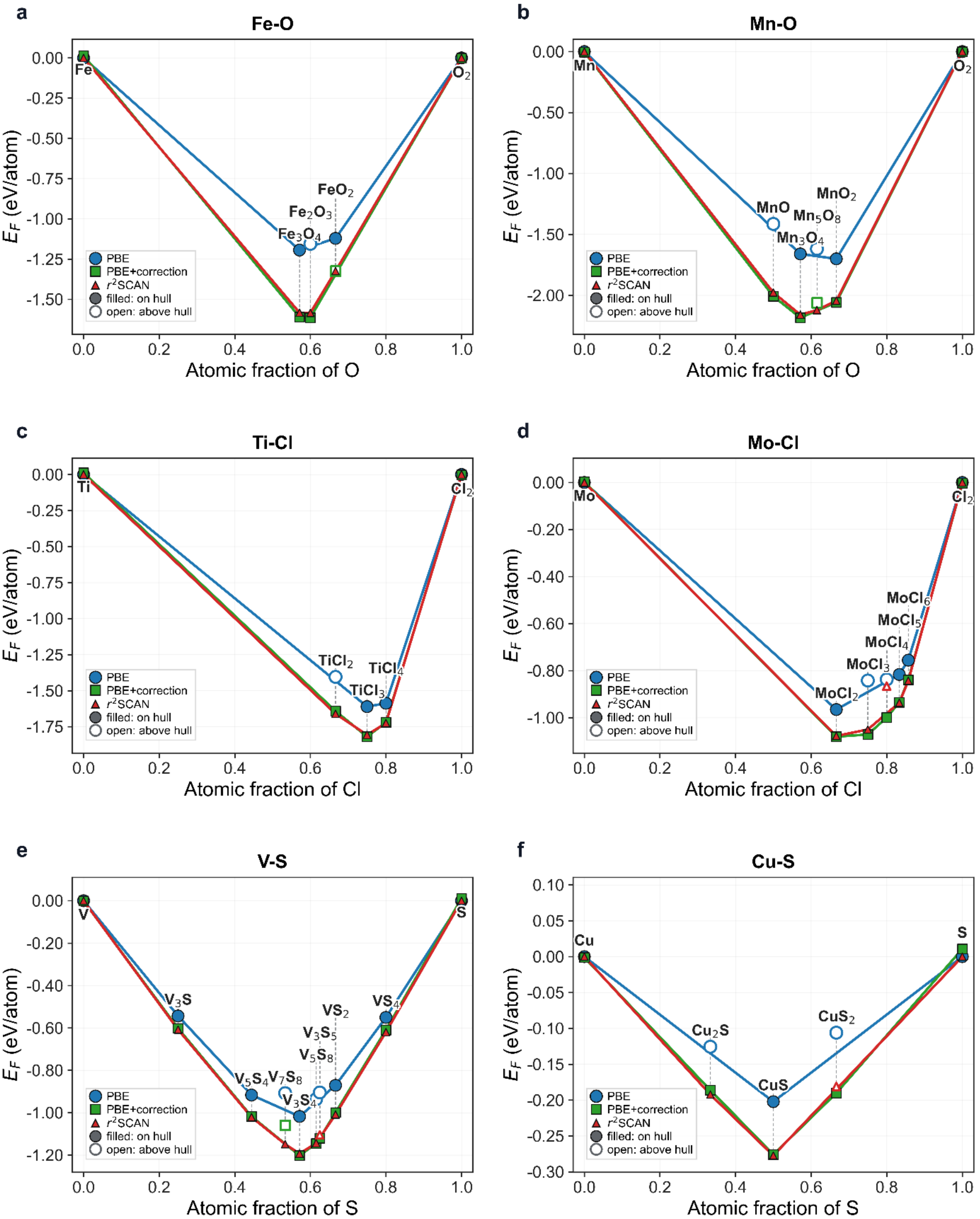

Figure 3. Comparison of binary phase stability obtained from PBE, PBE+correction, and r$^2$SCAN formation energies: (a) Fe-O, (b) Mn-O, (c) Ti-Cl, (d) Mo-Cl, (e) V-S, and (f) Cu-S. The plotted compositions are the union of convex-hull vertices identified by the three methods, and only the lowest-energy structure at each selected composition is shown for each energy scale. Filled symbols denote phases on the corresponding convex hull, whereas open symbols denote the lowest-energy structure at that composition lying above the hull. Blue circles, green squares, and red triangles represent PBE, PBE+correction, and r$^2$SCAN, respectively.

### 3.4 Battery voltage-profile prediction

To further test the model in a practical materials-property prediction task, we apply the correction to average Li deintercalation voltage calculations for Li-ion battery cathodes. Voltage plateaus are highly sensitive to relative energies among structures with different Li contents and therefore provide an important application case for evaluating formation-energy corrections.

We select $LiCoO_2$ and $LiMn_2O_4$ as representative systems. For each Li content, Li/vacancy configurations are generated using pymatgen, and the lowest-energy configuration is selected for voltage calculations. In $LiCoO_2$, PBE substantially underestimates the voltage relative to experiment and r$^2$SCAN, whereas the GPTFF-corrected voltage profile is nearly consistent with r$^2$SCAN and closer to experiment [33]. In $LiMn_2O_4$, PBE similarly underestimates the voltage plateaus, while r$^2$SCAN is closer to the experimental values [34]. The corrected model is broadly consistent with r$^2$SCAN and experiment, although it introduces an additional plateau transition at $x = 0.083$ and misses the transition at $x = 0.25$.

These results indicate that GPTFF correction can effectively reduce the systematic underestimation of PBE voltage profiles and make PBE formation energies more useful in electrochemical applications. The additional plateau in $LiMn_2O_4$ also highlights the extreme sensitivity of voltage profiles to small energy differences: local errors in configuration ordering may still affect phase-transition intervals and plateau shapes. More comprehensive Li/vacancy sampling and phase-stability analysis will therefore be needed to further validate model transferability in complex electrochemical systems.

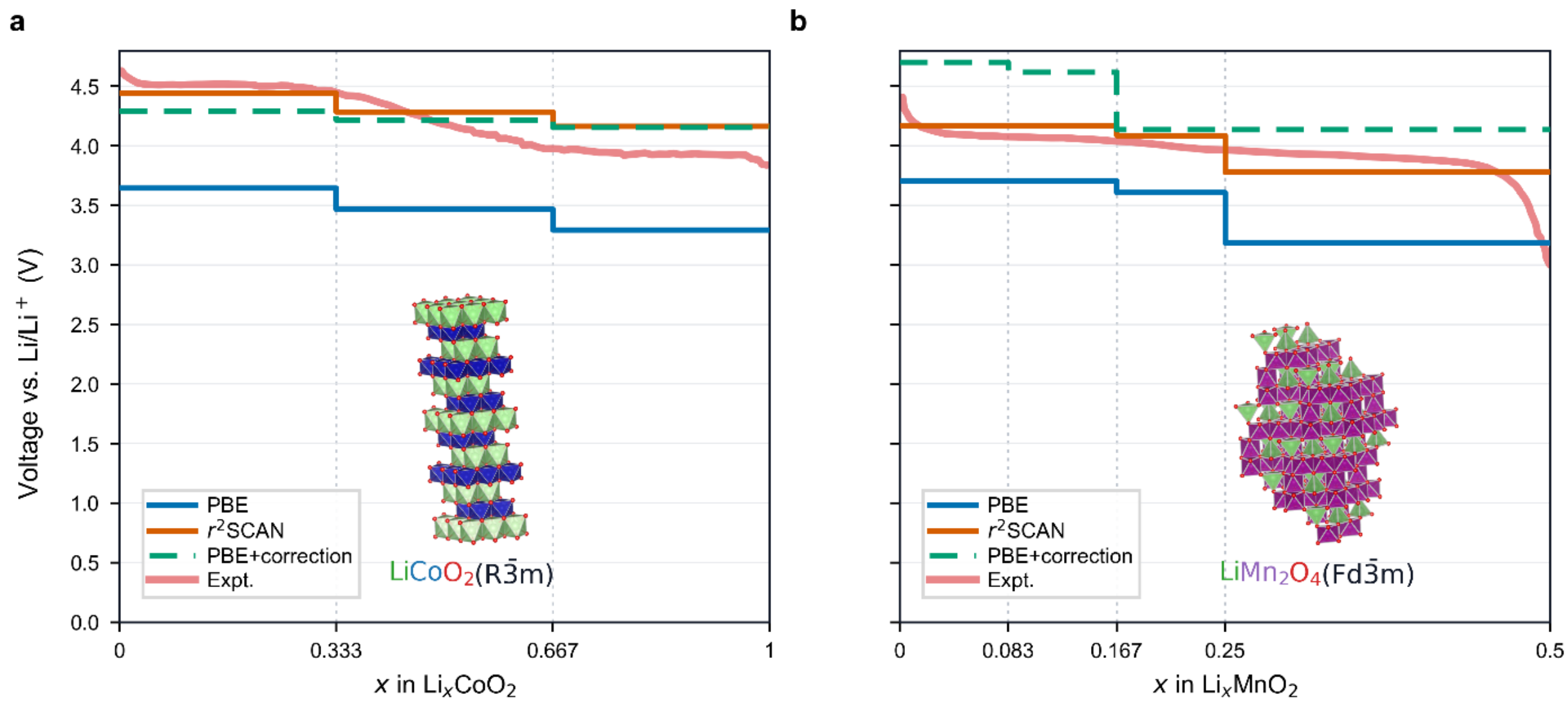

Figure 4. GPTFF-corrected formation energies applied to average Li deintercalation voltage prediction. (a) $LixCoO_2$. (b) $LixMnO_2$, normalized to MnO2 formula units. Blue, orange, green dashed, and pink curves represent PBE, r$^2$SCAN, PBE+correction, and experimental voltage plateaus, respectively.

## 3.5 Solid-state reaction pathways and reaction-energy prediction

Formation energies also directly determine the thermodynamic driving forces of chemical reactions. We therefore apply the model to representative solid-state reaction pathways and compare reaction enthalpies from PBE, $r^2$SCAN, model-corrected formation energies, and experiment [23].

For the $BaO + TiO_2$ reaction, PBE predicts $Ba_2TiO_4$ as the most stable product, whereas $r^2$SCAN and experiment both indicate that $BaTiO_3$ is more stable. After GPTFF correction, the model also predicts $BaTiO_3$ as the most stable product, with reaction energies consistent with $r^2$SCAN and experiment. For $Li_2O + Fe_2O_3$, PBE predicts $Li_3FeO_3$ as the most stable product, whereas $r^2$SCAN and experiment favor $LiFeO_2$. The corrected model also recovers $LiFeO_2$, although its reaction energy still differs from $r^2$SCAN and experiment by approximately 0.025 eV/atom.

For the $CuO + Fe$ and $MgCl_2 + CaO$ reactions, PBE, $r^2$SCAN, and experiment identify the same most stable products, and the corrected reaction energies move closer to $r^2$SCAN and experimental values. These results demonstrate that corrected formation energies can improve not only individual compound energies but also thermodynamic consistency in reaction networks involving multiple reactants and products.

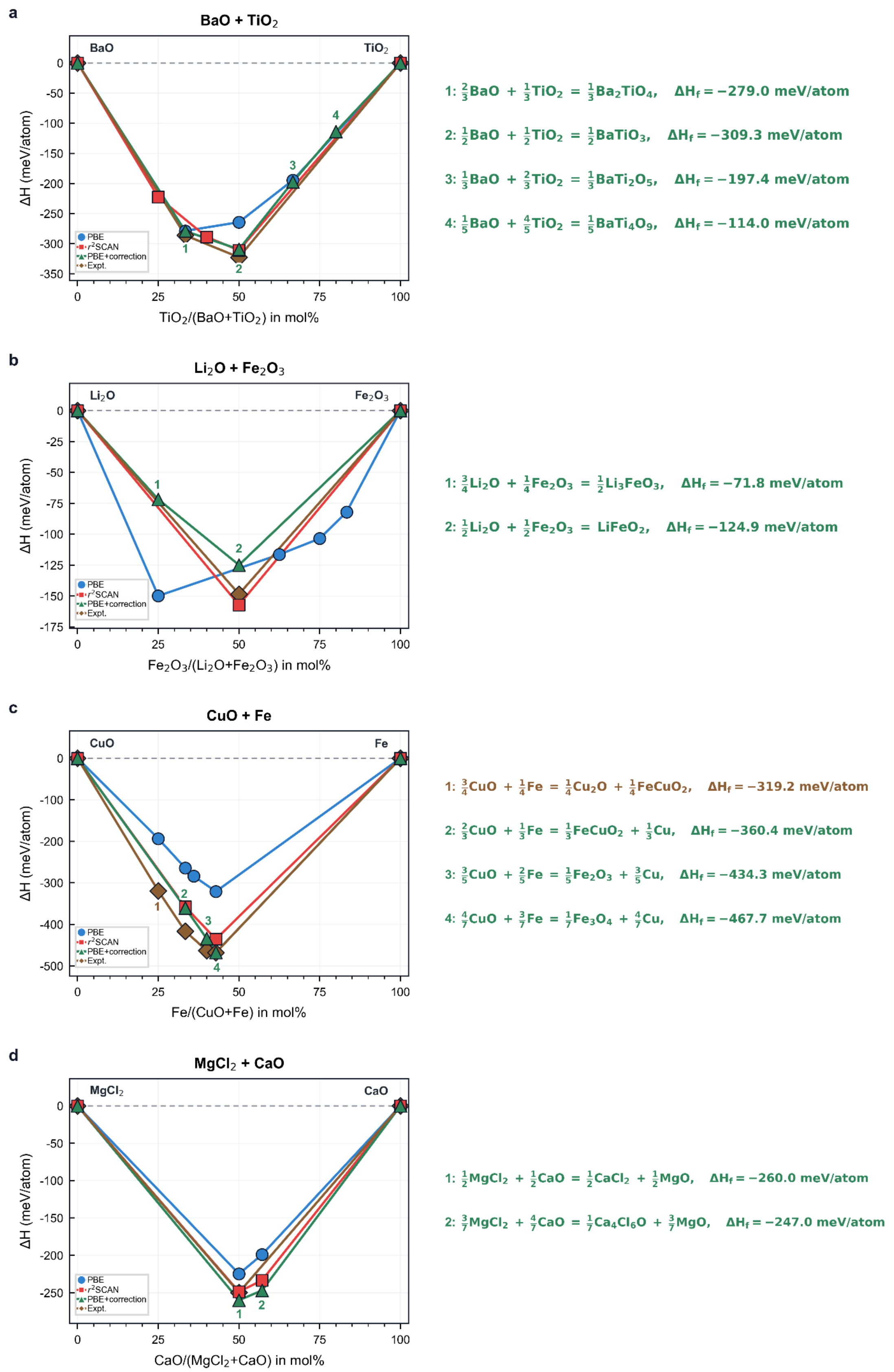

Figure 5. Comparison of reaction enthalpies along representative solid-state reaction pathways. (a) BaO-$TiO_2$; (b) Li2O-$Fe_2O_3$; (c) CuO-Fe; (d) $MgCl_2$-CaO. Star symbols mark the lowest-enthalpy products predicted by each method, and the dashed line indicates the reactant-endmember reference.

# 4 Discussion

Although this study verifies the strategy using PBE and $r^2$SCAN datasets as an example, the correction scheme is not bounded by specific DFT functionals or calculation parameters. This versatile pipeline is applicable to any pair of DFT datasets with differing computational configurations to realize consistent energy alignment.

Mainstream public materials databases including Materials Project [6], MatPES [21], OMat24 [13] and Alexandria [41] adopt inconsistent DFT setups, producing mismatched energy values that impede multi-source data fusion. With strong generalization ability, the proposed energy correction strategy can standardize energies across these databases, providing a universal tool for large-scale integration of computational materials data.

For future research, universal neural networks can be developed to automatically capture systematic energy deviations between various DFT datasets and realize fully automated cross-dataset energy alignment. Moreover, this energy correction strategy can be expanded to bridge DFT computational datasets and experimental databases, eliminating systematic energy inconsistencies between theoretical calculations and experimental measurements. Supplementing the training set with more paired computational and experimental samples will further improve the model's generalization and robustness. Such a framework delivers a promising and sustainable route for integrating multi-source theoretical and experimental materials data.

# 5 Conclusion

This work introduces a data-driven energy correction framework that unifies fragmented DFT datasets onto a shared energetic benchmark, outperforming rigid empirical correction methods in both accuracy and reliability. Trained on hundreds of thousands of structurally paired PBE–$r^2$SCAN crystal entries, our graph neural network seamlessly lifts low-cost PBE formation energies to $r^2$SCAN-grade thermodynamic fidelity, with rigorous tests on phase diagrams, battery voltage curves and solid-state reactions affirming its consistent real-world utility.

This residual-learning design is no narrow band-aid confined to a single pair of functionals. It readily standardizes misaligned energy values across all global public materials databases, and its core logic can stretch even further to bridge the persistent divide between computational DFT outputs and laboratory experimental results.

Beyond a mere technical tweak, this method tears down a long-standing barrier to large-scale multi-source materials data fusion. It paves an open, scalable path for the next generation of robust materials foundation models, and lights the way toward swifter, more insightful data-led discovery of transformative functional materials.

## Author Contributions

S.M. and M.L. conceived and supervised the project. Y.H. and J.H. performed dataset organization and model training. T.L. and H.K. supervised the model training process. Y.H. and M.L. drafted the original manuscript. All authors discussed the results and approved the final version. The authors declare no competing financial or personal interests that could have appeared to influence the work reported in this paper.

## Acknowledgements

This research is supported by National Key R&D Program of China (Grant No. 2024YFF0508500, 2021YFA1400200, and 2021YFA0718700), Chinese Academy of Sciences (Grant No. YSBR047), and National Natural Science Foundation of China (Grant No. 12450401).